\documentclass[amsmath,amssymb,showpacs,preprintnumbers,twocolumn]{revtex4}

\usepackage{graphicx}
\usepackage{dcolumn}
\usepackage{bm}

\graphicspath{{./pics/}}
\newcommand \be{\begin{equation}}
\newcommand \ee{\end{equation}}
\newcommand \ba{\begin{eqnarray}}
\newcommand \ea{\end{eqnarray}}

\begin{document}

\title{3D hybrid computations for streamer discharges and production of run-away electrons}
\author{Chao Li$^1$, Ute Ebert$^{1,2}$, Willem~Hundsdorfer$^{1,3}$}
\affiliation{$^1$Centrum Wiskunde and Informatica (CWI), P.O.Box 94079, 1090 GB Amsterdam, The Netherlands,}
\affiliation{$^2$Dept.\ Physics, Eindhoven Univ.\ Techn., The Netherlands,}
\affiliation{$^3$ Department of Science, Radboud University Nijmegen, The Netherlands.}


\begin{abstract}
We introduce a 3D hybrid model for streamer discharges that follows the dynamics of single electrons in the region with strong field enhancement at the streamer tip while approximating the many electrons in the streamer interior as densities. We explain the method and present first results for negative streamers in nitrogen. We focus on the high electron energies observed in the simulation. 
\end{abstract}


\maketitle

Streamers are a fundamental mode of electrical breakdown of ionizable matter when a strong voltage is applied; they are the first stage in  the evolution of sparks and lightning. Streamers are ionized plasma channels that grow into a non-ionized medium due to the self-enhancement of the electric field at their tips (see Fig.~\ref{fig:hy_result}). In this high field region, the electron energy distribution is very far from equilibrium and can have a long tail at high energies~\cite{Li2007,Li2008:1,Li2009:1}, which makes the streamer a plausible candidate for the generation of so-called run-away electrons~\cite{Dwy2005:1,Mos2006,Cha2008}. Such very energetic electrons subsequently can produce X-rays and $\gamma$-rays through Bremsstrahlung, therefore they may explain X-ray bursts and flashes observed during thunderstorms~\cite{Fis1994,Moo2001}, rocket triggered lightning~\cite{Dwy2003:2} and spark development in the laboratory~\cite{Dwy2005:2,Rahman2008,Ngu2008,Dwy2008}.

Streamer dynamics is mostly modeled by a fluid (or density) model~\cite{Vit1994,Kul1995,Liu2006,Mon2006:3,Luque2008:3,Pan2008} as this approximation is computationally most efficient and able to describe the main characteristics of the streamer discharge in a qualitative way. However, it obviously cannot trace the single particle dynamics. To follow the distribution of positions and velocities of individual electrons --- and therefore density fluctuations, run-away effects and excited molecular levels ---, a particle (or Monte Carlo) model~\cite{Bir1991,Kun1988:2,Cha2008} is required
that follows individual electrons and their elastic, inelastic and ionizing collisions with the background of abundant neutral molecules. However, the particle model is not suitable to study the single electron dynamics either, because the increasing number of electrons eventually renders computational power and storage unaffordable, while a super-particle approach causes numerical heating and stochastic artifacts~\cite{Li2008:2}.

We therefore here introduce a hybrid streamer model for full three-dimensional calculations. It uses the natural structure of the streamer: the particle model is applied in the most dynamic and exotic region with relatively few electrons and high local electric field, i.e., in the ionization front, and the many slow electrons inside the streamer channel are left to the fluid model. How to implement the spatial coupling of density and fluid model in the one-dimensional case, was presented in~\cite{Li2008:1,Li2009:1} and illustrated in Fig.~1 in both papers. Here the method is extended to 3D, and first results for a negative streamer in nitrogen at standard temperature and pressure are presented. We first discuss model and numerical implementation and then the physical results.

{\em Particle and extended fluid model and Poisson solver.} The particle and the fluid model were compared quantitatively in~\cite{Li2007,Li2009:1} for negative streamers in nitrogen for electric fields from $\sim$190 Td to $\sim$750 Td, or from 50 kV/cm to 200 kV/cm at standard temperature and pressure. An important finding in~\cite{Li2007} was that the electron and ion density in fluid or particle model start to differ when the field exceeds $\sim$190 Td, and, more importantly, that this relative difference largely increases with increasing electric field. In~\cite{Li2009:1}, the reasons of this density discrepancy are discussed, and it is shown that the fluid model had to be extended by a gradient expansion to optimally approximate the particle model. This model has the form
\ba \frac{\partial n_e}{\partial t} & + &  {\bf \nabla}\cdot {\bf j}_e = {\cal S},~~~ \frac{\partial n_p}{\partial t} = {\cal S}, \label{ch6_eq:fluid1and2}\\
{\bf j}_e &=& - \mu(E) {\bf E} n_e - {\bf D}({\bf E}) \cdot{\bf \nabla} n_e, \label{ch6_eq:flux} \\
{\cal S} & = &  \mu(E)\alpha(E)( E\; n_e + k_1(E){\bf E}\cdot\nabla n_e), \label{ch6_eq:Sl}
\ea
where $n_e$ and $n_p$ are electron and ion density, respectively, ${\bf j}_e$ is the electron flux and ${\cal S}$ is the nonlocal source term with a density gradient expansion parameterized by $k_1$, $\mu$ represents the mobility and ${\bf D}$ is the diffusion tensor, and ${\bf E}=(E_x, E_y, E_z)$ and $E$ are the electric field and its strength. The electric field is calculated with a fast Poisson solver: the 3D {\sc fishpack} subroutine~\cite{Sch1976,Bot1997}.

{\em Differential cross-sections.} The cross sections for the relevant collisions in the 3D particle model are taken from the {\sc siglo} database~\cite{Siglo} for incident electrons with energies up to 1 keV. Above 1~keV, the Born approximation~\cite{Liu1987} is used for elastic collisions, a fit formula in~\cite{Mur1988:1} is implemented for the electronically exciting collisions and the Born-Bethe approximation~\cite{Ino1971,Gar1988} is used for ionizing collisions. The electron transport coefficients, reaction rates and the average energies are generated in particle swarm experiments~\cite{Li2007,Li2009:1}; they agree well with the Boltzmann solver ({\sc bolsig+})~\cite{Siglo,Hag2005} when in both cases isotropic scattering and equal energy sharing in ionizing collisions is assumed. The scattering method derived by Okhrimovskyy {\it et al.}~\cite{Okh2002} is implemented for elastic and exciting collisions. Opal's empirical fit~\cite{Opa1971} is implemented for the energy splitting in ionizing events, where incident electrons with high energies are likely to keep most of their energy.

{\em The spatial coupling of fluid and particle model} --- more precisely, the position of the model interface as a function of the maximal field and the construction of the buffer region --- was already discussed in~\cite{Li2008:1,Li2009:1} for planar fronts. When the 3D streamer is decomposed into many narrow parallel columns oriented in the propagation direction (as detailed further below), this coupling can be applied in each of these columns. However, new problems arise due to the complexity of the 3D geometry: i) The model interface in 3D is never planar, but depending on the used criterion, it is either smoothly curved or even strongly fluctuating; a fluctuating model interface will create large buffer regions and dramatically increase the computational cost. Since in small grid cells, the electric field is smooth while the electron density can fluctuate heavily, the position of the model interface is determined here through the electric field rather than through the electron densities. More precisely, in the results shown below, the model interface in each column is placed where the field is $E=0.85~ E^+$ with $E^+$ being the maximal field ahead of the front within the column. This criterion ensures that the relative error for the electron densities in the streamer stays below 3\% for all fields E$^+$~\cite{Li2009:1}. The large region at the sides of the streamer that stays non-ionized, is treated by the particle model. ii) A direct contact of particle and fluid model without a buffer region can cause electron leaking, and hence loss of mass and charge. Therefore the buffer region has to be constructed carefully not only at the ionization front, but also in the lateral directions. Details on the model interface and the buffer region are given after introducing the structure of the simulation results.

\begin{figure}
 \begin{center}
 \includegraphics[width=.3\textwidth]{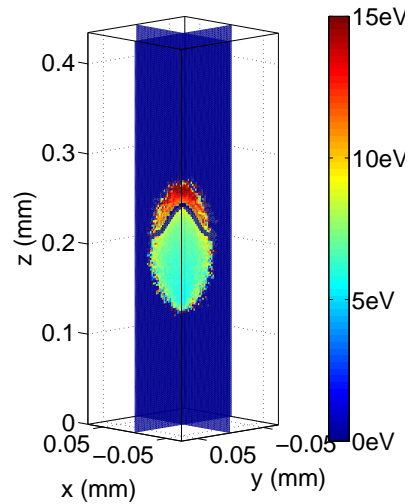}
\end{center}
\caption{Local mean energy of the electrons at time 0.32~ns when the simulation switches from pure particle to hybrid simulation; the buffer regions between particle and fluid regime are indicated with a dark curve line.}
     \label{fig:hy_en_split}
\end{figure}

\begin{figure*}
  \begin{center}
   \includegraphics[width=.98\textwidth]{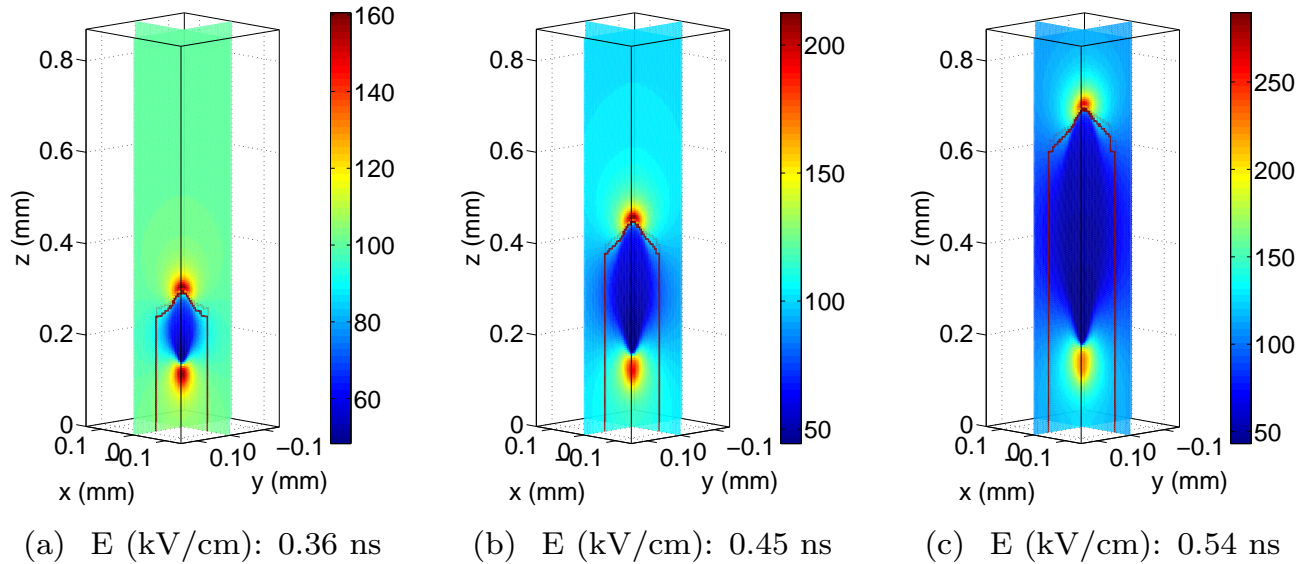}
\caption{\label{fig:hy_result} The electric field $E_z$ after time 0.36~ns, 0.45~ns and 0.54~ns of the simulation with model interface and buffer regions marked in red. All quantities are shown on two orthogonal planes that intersect with the 3D structure. The full system size is $0.59\times0.59\times1.18$~mm$^3$; only the dynamically interesting regions are shown.}
\end{center}
\end{figure*}

\begin{figure}
 \begin{center}
 \includegraphics[width=.3\textwidth]{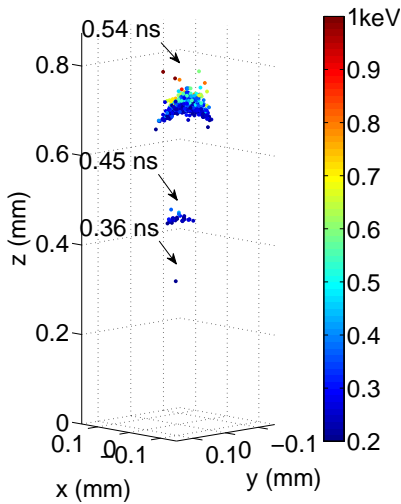}
\end{center}
\caption{The electrons with energy above 200 eV at time steps 0.36 ns, 0.45 ns and 0.54 ns. The maximal field $E^+$ at these times is 160, 220, and 290 kV/cm.
}
     \label{fig:hy_sca_runaway}
\end{figure}

\begin{figure}
 \begin{center}
 \includegraphics[width=.4\textwidth]{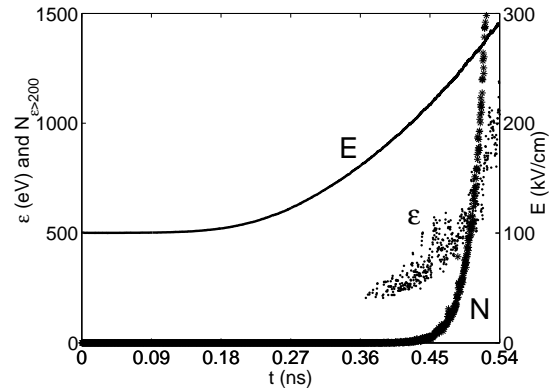}
\end{center}
\caption{The maximal electric field strength $E^+$, the number of electrons $N_{\epsilon>200~{\rm eV}}$ with energy larger than 200~eV and the highest electron energy $\epsilon_{max}$ as a function of time. The maximal electron energy exceeds 1~keV at t$\approx$0.54 ns.}
\label{fig:hy_runaway_ef_1}
\end{figure}

{\em A hybrid streamer simulation.} Figs.~\ref{fig:hy_en_split}--\ref{fig:hy_runaway_ef_1} show different aspects of the same simulation. It is a negative streamer in nitrogen at standard temperature and pressure. It propagates through a gap of 1.18~mm between two planar electrodes; the applied voltage is 11.8~kV which corresponds to a background field of 100~kV/cm or 372~Td. 
The simulation starts with 100 electrons and ions sitting 0.05 mm away from the cathode. They are initially followed by the pure particle model, and the hybrid model is introduced at time 0.32~ns when the number of electrons reaches $1.5\times10^7$
in a manner discussed further below.
The simulations are carried out on a uniform grid of $256 \times 256 \times 512$ grid points with the cell length $\Delta x= \Delta y =\Delta z=2.3~\mu$m and with time step $\Delta t= 0.3$~ps, the numerical procedure for particle and fluid model are described in~\cite{Li2007,Li2008:1,Li2009:1}. 

Fig.~\ref{fig:hy_en_split} shows the electron energy distribution at the moment when the simulation switches from pure particle to hybrid computations; the curved model interface is also marked. The figure shows that the region with high mean electron energies is covered by the particle model and the low energy part with many electrons is left for the fluid model; here the fluid model is both efficient and appropriate. Furthermore, the particle model is applied in all regions of low to vanishing electron density ahead and at the sides of the streamer; here the particle model is both more correct and also more efficient than the fluid model.

Fig.~\ref{fig:hy_result} shows the electric field $E_z$ in the direction of the background field at times 0.36~ns, 0.45~ns and 0.54~ns of the simulation (cf.~\cite{Mon2006:3,Ebe2006} for a more extended discussion and more plots of the streamer evolution in fluid approximation). The location of the buffer region is marked in red. The fluid model is applied within the red lines and the particle model is applied the large outer region where the field enhancement region is always included. As $E_z$ is large, the buffer region in $z$-direction should be 2 or even 3 cells long to obtain a stable electron flux at the model interface~\cite{Li2009:1}; this procedure is applied in each column where one column is one row of cells in the $z$ direction. In the $x$- and $y$-direction, one cell is long enough for the buffer region since the radial electric field is much smaller, but to prevent electron leaking from the particle region directly to the fluid region, 2 cells are used.

In practice, the hybrid simulation is very efficient in approximating the majority of the electrons by densities and in following the streamer much longer than the pure particle model. Specifically, at the times
0.36 ns, 0.45 ns and 0.54 ns shown in Fig.~\ref{fig:hy_result}, only  $12\%$, $7\%$ and $3\%$ of the electrons are followed individually. Nevertheless, the figure shows that in the region with the highest electric field the single electrons are followed. We remark that this simulation costs 43 hours on a normal desktop (Intel Quad2 CPU, 8 Gb RAM).

{\em Run-away electrons.}
The electron-nitrogen collision frequency is maximal for electron energies of about 200~eV, beyond that energy they have a chance to run away as the friction decreases when the energy increases further. Fig.~\ref{fig:hy_sca_runaway} therefore shows only the electrons with energy above 200~eV at the same three time steps as Fig.~\ref{fig:hy_result}. Electrons with $\epsilon>200$~eV start to appear when the maximal field $E^+$ reaches 160~kV/cm. But these electrons lose their energy almost immediately again. As the maximal field $E^+$ increases further during streamer propagation, both the number and the energy of the high energy electrons increases. Although most electrons still very quickly lose their energy, a few ones are able to accelerate further, and at time 0.54~ns, electrons with energy above 1~keV are observed. When the streamer later approaches the upper anode, the field increases further, also due to the proximity of the electrode, and electron energies up to 3.5 keV are seen.

Fig.~\ref{fig:hy_runaway_ef_1} analyzes the situation further. Plotted is the maximal electric field strength $E^+$, the number of electrons with energy above 200~eV, and the highest electron energy. Until approximately 0.2~ns, the maximal field equals the background field, i.e., the system is in the avalanche phase and no energetic electrons are present. After time 0.3~ns, the maximal field enhancement increases more than linearly in time, after 0.36~ns the first electrons above 200~eV appear, and after 0.45~ns their number and energies increase massively. Large fluctuations in the maximal electron energy as a function of time can be seen; there is not one electron that runs away, but many are being accelerated on average. Given the distance of 0.25mm that the front crosses between times 0.45~ns and 0.54~ns, the maximal electrostatic energy of the background field is $\approx 2.5$~keV; over this distance electrons accelerate from 0.2 to 1~keV.




During the time interval from 0.36~ns to 0.54~ns, the field at streamer head is enhanced to 1.5 to 3 times of the background field. These fields can accelerate electrons beyond the maximum of the electron-neutral friction force (cf. Fig.~2 in~\cite{Mos2006} or Fig.~9 in~\cite{Cha2008}) of 200 eV. 
Electrons in the energy range of several hundred eV get well ahead of the front, but many of them do not fully run away. As they get from the region of enhanced electric field to the region where the field decays ahead of the front while inelastic and ionizing scattering is still considerable, they are trapped and create many new small avalanches ahead of the ionization front~\cite{Gur1961,Kun1986:1,Bab2003}. 
The electrons with energies of several keV are likely to keep accelerating even in the lower background field ahead of the streamer~\cite{Bak2000,Mos2006}, but in the present simulation they rapidly reach the anode and disappear.

We have presented a 3D hybrid model for streamers that reliably can follow the single electron dynamics in the high field region of the streamer head at moderate computational costs, and that can observe electrons being accelerated to over 1 keV. Electrons with energies above 200~eV appear when the field enhancement at the streamer head exceeds 160~kV/cm or 600~Td. The energetic electrons can run out of the streamer head and relax somewhat ahead of the ionization front creating new avalanches; in this way they can create local front jumps and increase the mean velocity of the front. The investigation of streamers in air rather than in nitrogen will be subject of future studies, as well as the question whether streamers powered by higher voltages, e.g., in the corona of lightning leaders, can accelerate electrons into the relativistic range of MeV energies.

{\bf Acknowledgment:} The authors acknowledge the support of the Dutch National Program BSIK, in the ICT project BRICKS, theme MSV1. C.L. also acknowledges recent support through STW-project 10118 of the Netherlands' Organization for Scientific Research NWO.


\begin{thebibliography}{10}

\bibitem{Li2007}
Li C, Brok WJM, Ebert U, and van der Mullen JJAM 2007
\newblock {\em J. Appl. Phys.} {\bf 101} 123305

\bibitem{Li2008:1}
Li C, Brok WJM, Ebert U, and Hundsdorfer W 2008
\newblock {\em J. Phys. D: Appl. Phys.} {\bf 41} 032005

\bibitem{Li2009:1}
Li C, Ebert U, and Hundsdorfer W 2009
\newblock {\em submitted to J. Comput. Phys.}
\newblock (preprint available at arXiv:0904.2968)

\bibitem{Dwy2005:1}
Dwyer JR 2005
\newblock {\em Geophys. Res. Lett.} {\bf 32} L20808

\bibitem{Mos2006}
Moss GD, Pasko VP, Liu N, and Veronis G 2006
\newblock {\em J. Geophys. Res.} {\bf 111} A02307

\bibitem{Cha2008}
Chanrion O and Neubert T 2008
\newblock {\em J. Comput. Phys.} {\bf 227} 7222--7245

\bibitem{Fis1994}
Fishman GJ {\it et al.} 1994
\newblock {\em Science} {\bf 264} 1313

\bibitem{Moo2001}
Moore CB, Eack KB, Aulich GD, and Rison W 2001
\newblock {\em Geophys. Res. Lett.} {\bf 28} 2141--2144

\bibitem{Dwy2003:2}
Dwyer JR {\it et al.} 2003
\newblock {\em Science} {\bf 299} 694--697

\bibitem{Dwy2005:2}
Dwyer JR, Rassoul HK, and Saleh Z 2005
\newblock {\em Geophys. Res. Lett.} {\bf 35} L20809

\bibitem{Rahman2008}
Rahman M, Cooray V, Ahmad NA, Nyberg J, Rakov VA, and Sharma S 2008
\newblock {\em Geophys. Res. Lett.} {\bf 35} L06805

\bibitem{Ngu2008}
Nguyen CV, van Deursen APJ, and Ebert U. 2008
\newblock {\em J. Phys. D: Appl. Phys.} {\bf 41} 234012

\bibitem{Dwy2008}
Dwyer JR, Saleh Z, Rassoul HK, Concha D, Rahman M, Cooray V, Jerauld J, Uman MA, and Rakov VA 2008
\newblock {\em J. Geophys. Res.} {\bf 113} D23207

\bibitem{Vit1994}
Vitello PA, Penetrante BM, and Bardsley JN 1994
\newblock {\em Phys. Rev. E} {\bf 49} 5574--5589

\bibitem{Kul1995}
Kulikovsky AA 1995
\newblock {\em J. Phys. D: Appl. Phys.} {\bf 28} 2483--2493


\bibitem{Liu2006}
Liu N and Pasko VP 2006
\newblock {\em J. Phys. D: Appl. Phys.} {\bf 39} 327--334

\bibitem{Mon2006:3}
Montijn C, Hundsdorfer W, and Ebert U. 2006
\newblock {\em J. Comput. Phys.} {\bf 219} 801--835

\bibitem{Luque2008:3}
Luque A, Ratushnaya V, and Ebert U 2008
\newblock {\em J. Phys. D: Appl. Phys.} {\bf 41} 234005

\bibitem{Pan2008}
Pancheshnyi S, Segur P, Capeillere J, and Bourdon A 2008
\newblock {\em J. Comput. Phys.} {\bf 227} 6574--6590

\bibitem{Bir1991}
Birdsall CK and Langdon AB 1991
\newblock {\it Plasma Physics via Computer Simulation}.
\newblock (Adam Hilger)

\bibitem{Kun1988:2}
Kunhardt EE and Tzeng Y 1988
\newblock {\em Phys. Rev. A} {\bf 38} 1410--1421

\bibitem{Li2008:2}
Li C, Ebert U, and Brok WJM 2008
\newblock {\em IEEE Trans. on Plasma Science} {\bf 36} 914

\bibitem{Sch1976}
Schumann U and Sweet RA 1976
\newblock {\em J. Comput. Phys.} {\bf 20} 171--182

\bibitem{Bot1997}
Botta EFF, Dekker K, Notay Y, vander Ploeg A, Vuik C, Wubs FW, and  de~Zeeuw PM 1997
\newblock {\em Applied Numerical Mathematics} {\bf 24} 439--455

\bibitem{Siglo}
Morgan WL, Boeuf JP, and Pitchford LC 1995
\newblock {\em The \uppercase{Siglo} data base, \uppercase{CPAT} and \uppercase{K}inema software.}
\newblock {http://www.siglo-kinema.com}

\bibitem{Liu1987}
Liu JW 1987
\newblock {\em Phys. Rev. A} {\bf 35} 591--597

\bibitem{Mur1988:1}
Murphy T 1988
\newblock {\em Los Alamos National Lab. Report}

\bibitem{Ino1971}
Inokuti M 1971
\newblock {\em Rev. Mod. Phys.} {\bf 43} 297--347

\bibitem{Gar1988}
Garcia G, Perez A, and Campos J 1988
\newblock {\em Phys. Rev. A} {\bf 38} 654--657

\bibitem{Hag2005}
Hagelaar GJM and Pitchford LC 2005
\newblock {\em Plasma Sources Sci. Technol.} {\bf 14} 722--733

\bibitem{Okh2002}
Okhrimovskyy A, Bogaerts A, and Gijbels R 2002
\newblock {\em Phys. Rev. E} {\bf 65} 037402

\bibitem{Opa1971}
Opal CB, Peterson WK, and Beaty EC 1971
\newblock {\em The Journal of Chemical Physics} {\bf 55} 4100--4106

\bibitem{Ebe2006}
Ebert U, Montijn C, Briels TMP, Hundsdorfer W, Meulenbroek B, Rocco A, and van Veldhuizen EM 2006
\newblock {\em Plasma Sources Sci. Technol.} {\bf 15} S118

\bibitem{Gur1961}
Gurevich AV 1961
\newblock {\em Sov. Phys. JETP} {\bf 12} 904--912

\bibitem{Kun1986:1}
Kunhardt EE and Tzeng Y 1986
\newblock {\em Phys. Rev. A} {\bf 34} 2158--2166

\bibitem{Bab2003}
Babich LP 2003
\newblock {\em High-energy phenomena in electric discharges in dense gases:
  theory, experiment and natural phenomena}.
\newblock (Futurepast, Arlington, Virginia)

\bibitem{Bak2000}
Bakhov KI, Babich LP and Kutsyk IM
\newblock {\em IEEE Trans. on Plasma Science} {\bf 28} 1254


\end{thebibliography}


\end{document}